# Linking dwarf galaxies to halo building blocks with the most metal-poor star in Sculptor[*]

Anna Frebel[1], Evan Kirby[2,3], and Joshua D. Simon[4]

[*]*This paper includes data gathered with the 6.5 meter Magellan Telescopes located at Las Campanas Observatory, Chile.*

[1]*Harvard-Smithsonian Center for Astrophysics, Cambridge, MA 02138, USA; afrebel@cfa.harvard.edu*

[2]*California Institute of Technology, Pasadena, CA 91106, USA; enk@caltech.edu*

[3]*Hubble Fellow*

[4]*Observatories of the Carnegie Institution of Washington, Pasadena, CA 91101, USA; jsimon@ociw.edu*

**Current cosmological models[1,2] indicate that the Milky Way's stellar halo was assembled from many smaller systems. Based on the apparent absence of the most metal-poor stars in present-day dwarf galaxies, recent studies[3] claimed that the true Galactic building blocks must have been vastly different from the surviving dwarfs. The discovery of an extremely iron-poor star (S1020549) in the Sculptor dwarf galaxy based on a medium-resolution spectrum[4] cast some doubt on this conclusion. However, verification of the iron-deficiency and measurements of additional elements, such as the α-element Mg, are mandatory for demonstrating that the same type of stars produced the metals found in dwarf galaxies and the**

**Galactic halo. Only then can dwarf galaxy stars be conclusively linked to early stellar halo assembly. Here we report high-resolution spectroscopic abundances for 11 elements in S1020549, confirming the iron abundance of less than 1/4000th that of the Sun, and showing that the overall abundance pattern mirrors that seen in low-metallicity halo stars, including the α-elements. Such chemical similarity indicates that the systems destroyed to form the halo billions of years ago were not fundamentally different from the progenitors of present-day dwarfs, and suggests that the early chemical enrichment of all galaxies may be nearly identical.**

Previous attempts[3] to uncover the chemically primitive stellar populations in the relatively luminous ($L>10^5$ $L_{Sun}$) classical dwarf galaxies suggested a statistically significant absence of extremely metal-poor stars with [Fe/H]<−3.0 (where [A/B]=$\log(N_A/N_B)$−$\log(N_A/N_B)_{Sun}$ for the number N of atoms of elements A and B). The higher metallicity stars in those systems also had abundance patterns different from those of equivalent halo stars,[5,6] indicating different enrichment timescales for the classical dwarfs, at least during their later evolution. Given these discrepancies, it was suggested that dwarf galaxies do not fit[7] into the hierarchical merging model[8], and objects related to the present-day dwarf galaxies could be excluded from making a significant contribution to the metal-poor halo. The apparent lack of extremely metal-poor stars may have stemmed either from the small samples employed by high-resolution spectroscopic searches or reliance on the near-infrared Ca triplet lines to infer Fe abundances indirectly from low-resolution spectra. By revisiting the chemical inventory of these systems with new techniques[9] one such extremely metal-poor star was recently discovered[4], and we report its detailed chemical composition here.

On 2009 July 26-27, we obtained a high-resolution spectrum of the faint (apparent visual magnitude V=18.2 mag) red giant star S1020549 located in the Sculptor dwarf spheroidal galaxy. Using the MIKE spectrograph[10] at the Magellan-Clay telescope, we employed a 0.7 arcsec slit width, yielding a spectral resolution of R~36,000 at blue wavelengths, and R~30,000 at red wavelengths. Figure 1 shows a portion of the spectrum of this star that demonstrates its low metal content. The radial velocity of 118.6±0.2 km/s is in good agreement with the systemic velocity of Sculptor of 110.6 km/s (with a dispersion of 10.1 km/s; ref 11). This confirms the Sculptor membership of our target. Stellar parameters, abundances and uncertainties were determined using the same techniques outlined in ref. 12. We derive an effective temperature of $T_{eff}$=4,550 K, a surface gravity of log g=0.9, a metallicity of [Fe/H]=−3.81, and a micro-turbulent velocity of $v_{micr}$=2.8 km/s.

Chemical abundances were obtained for a total of eight elements and upper limits were measured for three others (see Table 1). The most outstanding feature is the extremely low Fe abundance, confirming the initial value estimated.[4] With [Fe/H]~−3.8, it has the lowest Fe abundance of all known stars in any of the dwarf galaxies. This is at variance with earlier claims[3] that the relatively luminous dwarfs such as Sculptor do not host stars with [Fe/H]≤−3.0. S1020549 is almost a factor of 6 less metal-enriched than this value.

Generally, the elemental ratios suggest that this star has a chemical pattern nearly identical to that of similarly metal-poor halo stars. Figure 2 shows the abundances of S1020549 in comparison with Galactic halo and dwarf galaxy stars collected from the literature. Moreover, the measurements of Mg, Ca, Sc, Ti, and Cr agree well with the trends found in halo stars below [Fe/H]~−3.5, while mostly disagreeing with the values of the more metal-rich stars in the classical dwarf galaxies. Si was difficult to measure but indicates a similar behaviour as in the halo. The C and Na abundances are also consistent with halo material, but the large scatter in the halo data (~1.0 dex for Na and ~2 dex for C below [Fe/H]=−3.0) precludes strong conclusions. Nevertheless, it is interesting that S1020549 is not strongly carbon-enhanced relative to iron (the upper limit being [C/Fe]<0.15) as many of the halo stars with [Fe/H]<−3.5 have values of [C/Fe]>1.0. However, the archetypical halo star CD −38 245 at [Fe/H]~−4.0 has an upper limit[13] of [C/Fe]<−0.3, illustrating the large carbon variations at the lowest metallicities. Among the comparison stars from the halo, a large scatter of up to ~3 dex is also observed for neutron-capture elements such as Sr and Ba. The upper limits for these two elements indicate significant deficiencies in S1020549 but they are still consistent with the most neutron-capture depleted halo stars (see Figure 2). Extremely low neutron-capture levels generally seem to be a typical signature of metal-poor stars in both the classical dwarfs[14] as well as the much less luminous ultra-faint dwarf galaxies[12,15] even at higher metallicities. The overall abundance pattern of S1020549 suggests that the early evolution of Sculptor was dominated by a population of massive stars that were responsible for the production of α-elements while contributing few of the heaviest elements.

A bias-corrected metallicity distribution function (MDF) for the Milky Way halo was recently determined from medium-resolution spectra of metal-poor stars found in the Hamburg/ESO survey.[16] The MDF drops sharply at [Fe/H]~−3.6, pointing to the extreme rarity of the lowest metallicity stars in the halo. In fact, only 9 halo stars are currently known with metallicities lower than that of S1020549, including the star with the lowest Fe abundance, HE1327−2327[17] with [Fe/H]=−5.4.

In contrast, the MDFs of the ultra-faint dwarfs suggest that these tiny systems contain a larger fraction of extremely metal-poor stars than the Milky Way halo does.[18] However, because only a small number of stars have been observed in those galaxies, no stars nearly as metal-poor as S1020549 have yet been identified there. Thus, even though a similar overlap in abundances between halo objects and six stars in two ultra-faint dwarfs was found from a recent high-resolution spectroscopic analysis[12], the objects in the two ultra-faint dwarf galaxies are factors of 3-25 more metal-rich than S1020549. Given the absence of extremely metal-poor stars in the classical dwarfs until now, it is currently unclear how the metal-poor tail of the MDF of these systems compares with that of the halo and the ultra-faint dwarfs. In any case, S1020549 is a very rare object and likely closely tied to the formation of the very first stars in dwarf galaxies.

The presence of a star with [Fe/H]=−3.81 in a relatively modest survey[4] indicates that future observational searches should discover more such objects in Sculptor and other dwarf galaxies. But collecting the required high-resolution data for these distant, and hence very faint, member stars in any of the dwarfs challenges currently available 6-10 meter telescopes. The next generation of extremely large optical telescopes, once equipped with high-resolution spectrographs, will thus open up a whole new window for thoroughly studying early galaxy assembly through stellar chemistry.

The agreement of the abundance ratios of S1020549 with those of the halo suggests that the classical dwarf spheroidal galaxies underwent very similar chemical enrichment in the earliest phases to the MW halo and ultra-faint dwarfs. In particular, this result provides evidence that the early chemical evolution of galaxies spanning a factor of more than 1 million in luminosity is dominated by the same type of stars, and possibly the same mass function, namely massive core-collapse supernovae as indicated by the α-element enhancement found in all of the most metal-poor stars. This universality would also characterize any dwarf galaxies that were accreted at early times to build the Milky Way and its stellar halo. Those accreted systems are therefore unlikely to have been significantly different from the progenitors of the surviving dwarfs. In that case, the oldest, most metal-poor stars observed in present-day dwarf galaxies should be representative of the stars found in the Galactic building blocks before their destruction. The latest observational results thus support the idea that mergers and accretion of small, generally metal-poor systems, as predicted by ΛCDM models, can in principle explain the metal-poor stellar content of the Galactic halo. Nevertheless, we cannot rule out the possibility that some metal-poor stars in the halo came from more massive systems than the ones considered here. Currently, the most metal-poor stars known in the Milky Way's most luminous satellites have [Fe/H]>∼−2; future discoveries of extremely metal-poor stars in these galaxies would be required to demonstrate the feasibility of this new picture.

The existence of S1020549 is not only interesting in itself, but is of particular importance because the star resides in a satellite of the Milky Way located at a distance of about 86 kpc.[19] It has become clear that the Galactic halo itself is complex, with at least two populations of different origins. The outer halo (beyond 15-20 kpc from the Galactic centre) has an average metallicity a factor of four lower than that of the inner halo.[20] This is in line with many of the most metal-poor halo stars showing some kinematic properties characteristic of the outer halo (e.g., large [retrograde] radial velocities, or large space velocity components).

The picture that is now emerging from the abundance agreement found in the present study (see Fig. 2) thus suggests that the old, low-metallicity tail of the outer halo could have been populated with metal-poor stars deposited by small dwarf galaxies that were destroyed long ago. But were enough dwarf galaxies accreted to account for all of the metal-poor halo stars? The surviving ultra-faint dwarfs are the least luminous and most dark matter-dominated galaxies,[21] and they possess very few stars despite containing some extremely metal-poor stars. It is thus unclear whether the accretion of even large numbers of analogues to such systems can provide enough stellar mass to account for the entire population of low-metallicity field stars. On the other hand, massive satellites like the progenitors of the Magellanic Clouds are thought to have provided the vast majority of the inner halo.[22,23] Galaxies with stellar masses in between massive gas-rich objects (early versions of today's Magellanic Clouds) and less luminous systems (appearing today as ultra-faint dwarfs), hence appear to be more natural candidates for providing the metal-poor stellar content of the outer halo.

We thank Lars Hernquist for discussions on galaxy formation. A. F. acknowledges support through a Clay Fellowship administered by the Smithsonian Astrophysical Observatory. Support for this work was provided by NASA through a Hubble fellowship grant awarded by the Space Telescope Science Institute, which is operated by the Association of Universities for Research in Astronomy, Inc, for NASA (for E. N. K.). J. D. S. acknowledges the support of a Vera Rubin Fellowship provided by the Carnegie Institution of Washington.

The authors declare that they have no competing financial interest.

Correspondence and requests for materials should be addressed to A. F. (e-mail: afrebel@cfa.harvard.edu).

Author Contributions A.F. took the high-resolution observations, led the analysis and the write-up of the paper; E.K. provided the target and contributed to the text writing; J.S. contributed to the analysis and text writing.

Figure 1 caption: Spectral comparison of S1020549 with two other metal-poor stars. The high-resolution spectra show the Mg b triplet lines at 5170A.

Information on the stellar parameters is given for each star. All three stars have similar effective temperatures. The difference in line strengths reflects the different metallicities, illustrating that S102549 is much more heavy-element deficient than the two other objects (which are located in the ultra-faint dwarf galaxies Ursa Major II and Coma Berenices).[12] The total exposure time of the data collected for S1020549 was 7.55 h. The Carnegie-Python pipeline (D. Kelson 2009, priv. comm.) was used for the echelle data reduction; individual frames from each night were combined and reduced, and then added into the final spectrum. The signal-to-noise ratio of the binned spectrum is S/N~22 per 66 mÅ pixel at ~460 nm and S/N~56 per 133 mÅ pixel at ~640nm. The spectrum immediately revealed the metal-poor nature of this object, which is among the faintest metal-poor stars for which a high-resolution spectrum has ever been obtained. S1020549 (coordinates right ascension, RA=01h 00m 47.8s and declination=−33 41 03 at Equinox 2000) was discovered from employing a new measurement technique[4] that uses actual Fe lines present in medium-resolution (R~6000) spectra to obtain robust estimates for the Fe abundances of large numbers of stars. This technique already led to the discovery of two extremely metal-poor stars in ultra-faint dwarf galaxies[12] and additional candidates in need of high-resolution follow-up[4,18,24]. The Fe abundances of these two objects were confirmed from high-resolution spectroscopy[11] which also provided the required data quality to obtain detailed measurements for many additional elements.

Figure 2 caption: Abundance ratios as a function of iron abundance in S1020549 and other metal-poor stars. In eight elements, S1020549 (big green filled circle at [Fe/H]~−3.8) is compared with halo stars (black circles; refs. 13, 25, 26, 27), ultra-faint dwarf galaxy stars (blue diamonds; ref. 12), and the brighter dwarf galaxy stars (pink and yellow diamonds; refs. 5, 6, 14, 15, 28, 29). Small green circles indicate higher metallicity Sculptor targets. The abundances ([X/Fe]) were derived from a 1D LTE plane-parallel model atmosphere analysis using a Kurucz model (http://kurucz.harvard.edu). Error bars refer to the standard error of the abundances obtained from the individual lines for each element, but systematic uncertainties presented in Table 1 should also be considered. For Na, we present abundances that are not corrected for effects of non-local thermodynamic equilibrium (LTE). Other elements could also be affected by departures from LTE. Examples are discussed in ref. 13. However, the comparison with the LTE abundances of all halo stars would not be significantly affected, if at all.

**Chemical abundances of S1020549**

| Species | Log ε(X) | [Element/H] | [Element/Fe] | $N_{lin}$ | $\sigma_{random}$ | $\sigma_{syst.}$ |
|---------|----------|-------------|--------------|-----------|----------|---------|
| C (CH) | <4.78 | <−3.65 | <0.16 | 1 | … | … |
| NaI | 2.31 | −3.93 | −0.12 | 2 | 0.07 | 0.22 |
| Si | 4.66: | −2.85: | 0.96: | 1 | 0.30 | 0.39 |
| MgI | 3.90 | −3.70 | 0.11 | 2 | 0.25 | 0.23 |
| CaI | 2.80 | −3.54 | 0.27 | 4 | 0.15 | 0.22 |
| ScII | −0.51 | −3.66 | 0.15 | 5 | 0.20 | 0.16 |
| TiI | 1.38 | −3.57 | 0.24 | 4 | 0.06 | 0.29 |
| TiII | 1.29 | −3.66 | 0.15 | 16 | 0.21 | 0.14 |
| CrI | 1.23 | −4.41 | −0.60 | 3 | 0.12 | 0.25 |
| FeI | 3.69 | −3.81 | … | 63 | 0.18 | 0.25 |
| FeII | 3.66 | −3.84 | −0.03 | 4 | 0.23 | 0.14 |
| SrII | <−2.70 | <−5.57 | <−1.76 | 1 | … | … |
| BaII | <−3.10 | <−5.28 | <−1.47 | 1 | … | … |

Table comment: Log $\varepsilon(X)$ refers to log(Element/H)+12; [Element/Fe] ratios are computed using the [FeI/H] abundance and the solar abundances given in ref. 30, $N_{lin}$ lists the number of employed elemental lines. A colon denotes a difficult measurement due to the absorption line being located at a wavelength of ~3900Å where is signal-to-noise of the data is low. $\sigma_{random}$ is the standard error of the abundances obtained from individual line measurements for each element. For elements with just one line, a nominal value of 0.3 dex has been adopted. $\sigma_{total}$ are the quadratic sums of all individual systematic uncertainties which were obtained from varying one stellar parameter at a time by its uncertainty ($\Delta T_{eff}$=200 K, $\Delta$log g=0.4 dex, $\Delta v_{micr}$=0.3 km/s) and an assessment of typical continuum placement uncertainties ranging from 0.1 to 0.2 dex.

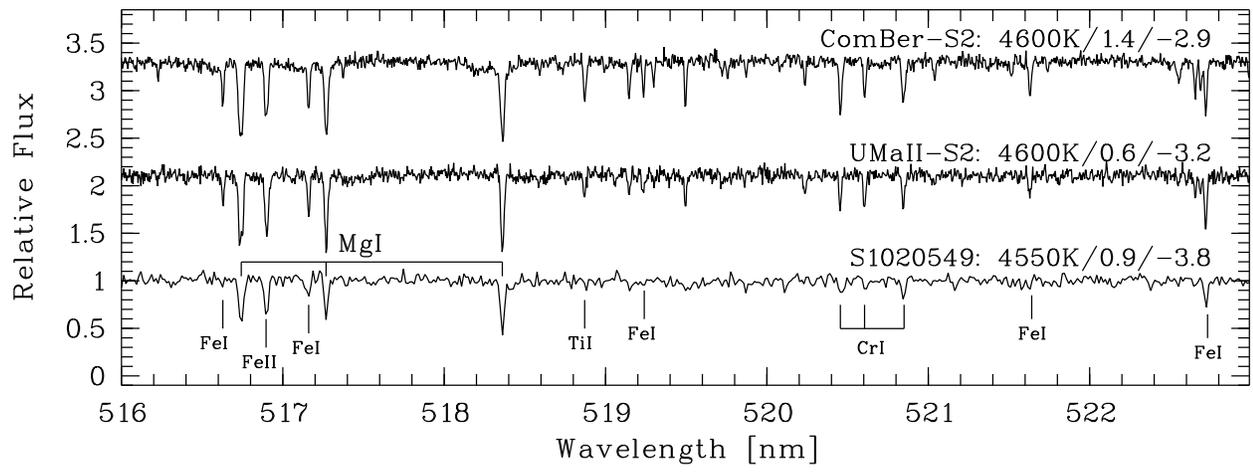

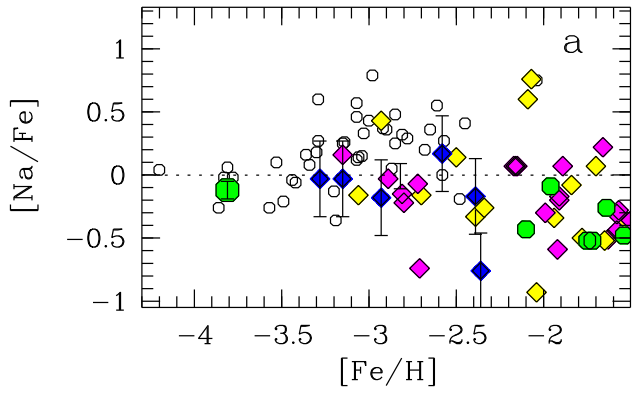
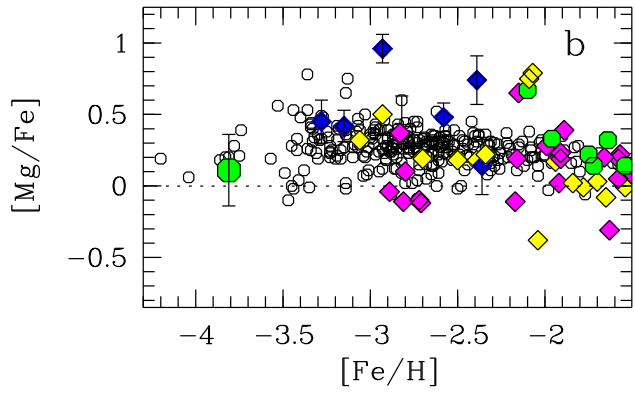
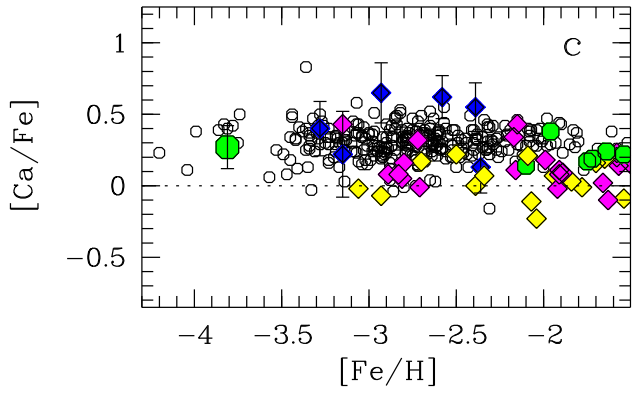
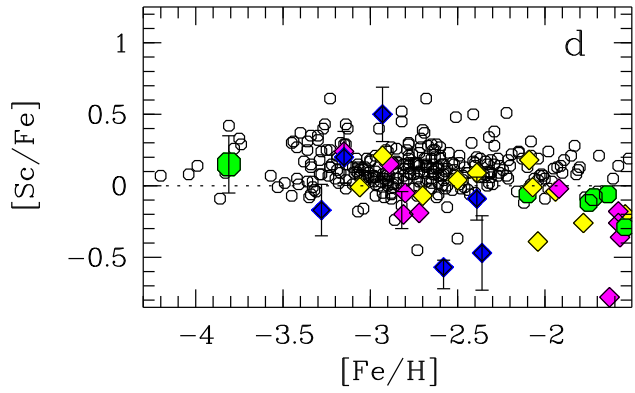
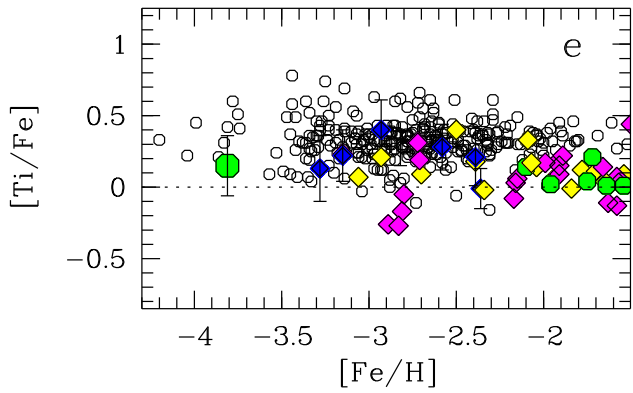
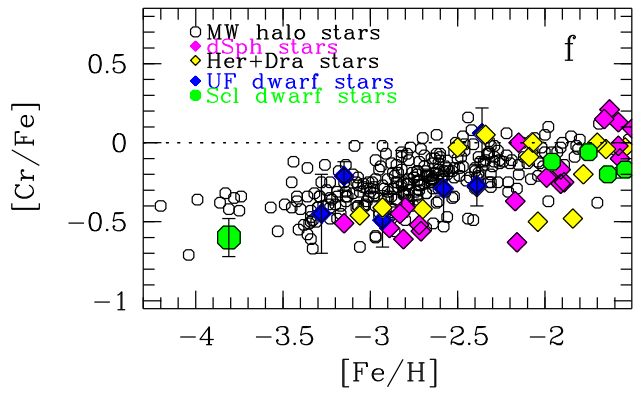
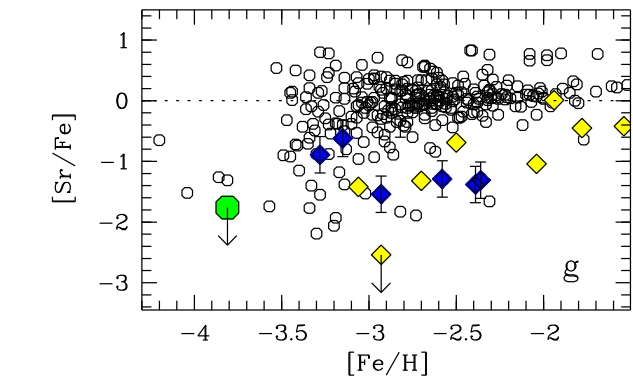
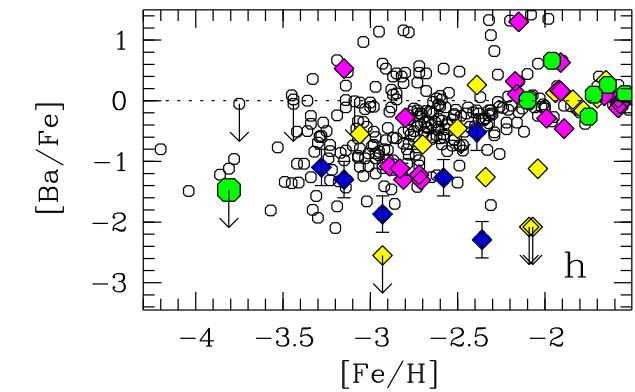